\documentclass{article}

\usepackage{PRIMEarxiv}
\usepackage{fancyhdr}
\usepackage[utf8]{inputenc} 
\usepackage[T1]{fontenc}    
\usepackage{hyperref}       
\usepackage{url}            
\usepackage{booktabs}       
\usepackage{amsfonts}       
\usepackage{nicefrac}       
\usepackage{microtype}      
\usepackage{lipsum}
\usepackage{fancyhdr}       
\usepackage{graphicx}       
\graphicspath{{media/}}     

\title{IoDResearch: Deep Research on Private Heterogeneous Data via the Internet of Data}
\author{
  Zhuofan Shi, Zijie Guo, Xinjian Ma, Gang Huang, Yun Ma, Xiang Jing\thanks{Corresponding author: jingxiang@pku.edu.cn \\ © 2025 IEEE. Personal use of this material is permitted. Permission from IEEE must be obtained for all other uses, in any current or future media, including reprinting/republishing this material for advertising or promotional purposes, creating new collective works, for resale or redistribution to servers or lists, or reuse of any copyrighted component of this work in other works.} \\
  Peking University \\
  National Key Laboratory of Data Space Technology and System \\
}

\begin{document}
\maketitle

\begin{abstract}
The rapid growth of multi-source, heterogeneous, and multimodal scientific data has increasingly exposed the limitations of traditional data management. Most existing Deep Research (DR) efforts focus primarily on web search while overlooking local private data. Consequently, these frameworks exhibit low retrieval efficiency for private data and fail to comply with the FAIR principles, ultimately resulting in inefficiency and limited reusability.
To this end, we propose IoDResearch (Internet of Data Research), a private data-centric Deep Research framework that operationalizes the Internet of Data paradigm. IoDResearch encapsulates heterogeneous resources as FAIR-compliant digital objects, and further refines them into atomic knowledge units and knowledge graphs, forming a heterogeneous graph index for multi-granularity retrieval. On top of this representation, a multi-agent system supports both reliable question answering and structured scientific report generation. 
Furthermore, we establish the IoD Deep Research Benchmark to systematically evaluate both data representation and Deep Research capabilities in IoD scenarios. Experimental results on retrieval, question answering, and report-writing tasks show that IoDResearch consistently surpasses representative retrieval-augmented generation (RAG) and Deep Research baselines. Overall, IoDResearch demonstrates the feasibility of private-data-centric Deep Research under the IoD paradigm, paving the way toward more trustworthy, reusable, and automated scientific discovery. URL: \url{https://github.com/FredericVAN/PKU\_IoDResearch}
\end{abstract}

\keywords{Internet of Data \and Deep Research \and LLM}

\section{Introduction}
\label{sec:intro}
In the data-driven era of artificial intelligence, the limitations of traditional data management approaches have become increasingly evident with the explosive growth of data, making the efficiently managing and leveraging massive heterogeneous data has emerged as a pressing challenge
\cite{wang2025towards,putrama2024heterogeneous}, especially in scientific research scenarios that require robust cross-domain data infrastructures \cite{jingru2024technical}.Because Real-world scientific data are often multi-source, heterogeneous, and multimodal, lacking unified representation standards and interoperability mechanisms, which leads to low efficiency in data utilization. For example, materials science research typically involves vast amounts of heterogeneous data derived from experiments, simulations, and literature records \cite{Data-Driven-Materials-Science}. Effective management, sharing, and analysis of these data are essential for accelerating scientific discovery and fostering innovation. However, due to inherent differences in data formats, structures, and semantics, as well as the ever-increasing scale of data, traditional centralized management approaches face severe challenges \cite{Knowledge-Graph-Empowered-Materials-Discovery}.

To address these limitations, the Internet of Data (IoD)\cite{shujukongjian,zhang2023identifier} has emerged as a data-centric paradigm designed to overcome the limitations of traditional data management. IoD leverages open, data-centric software architectures and standardized interoperability protocols to interconnect heterogeneous platforms and systems, thereby forming a unified data network. Rooted in the Digital Object Architecture (DOA)\cite{kahn2006framework}, IoD provides persistent identifiers, enriched metadata, and standardized interoperability mechanisms to encapsulate diverse resources as digital objects (DOs).

In parallel, the FAIR principles, proposed by Wilkinson et al.\cite{wilkinson2016fair}, were established to address the lack of common standards in the face of rapidly growing scientific data. They ensure that data are Findable, Accessible, Interoperable, and Reusable. Without these four dimensions, data would remain locked in isolated systems, making cross-domain retrieval and reuse extremely difficult. By design, IoD is inherently aligned with the FAIR principles. Through these mechanisms, IoD effectively eliminates data silos, enabling seamless linking and reuse of distributed heterogeneous data, and thus lays a solid foundation for cross-domain scientific data infrastructures.

More recently, the rapid advancement of large language models has accelerated the development of Deep Research (DR) agents\cite{huang2025deepresearchagentssystematic} has gained significant attention in both academia and industry. Such agents exhibit a wide range of promising abilities, including proposing novel research directions \cite{si2024canllm, hu2024nova}, efficiently acquiring information through search-augmented tools \cite{Search-o1, jin2025search}, and performing preliminary analyses or experiments before composing full research reports or papers \cite{ai-scientist-v2, zheng-etal-2024-openresearcher}.

However, existing DR systems still rely predominantly on web search, with limited exploration of local heterogeneous and multimodal data. Consequently, current approaches largely overlook these data characteristics and often fail to comply with the FAIR principles \cite{wilkinson2016fair}. Such limitations restrict the effective circulation and reuse of private data resources and impede the discovery of latent relationships.

\begin{figure}
    \centering
    \includegraphics[width=1.0\linewidth]{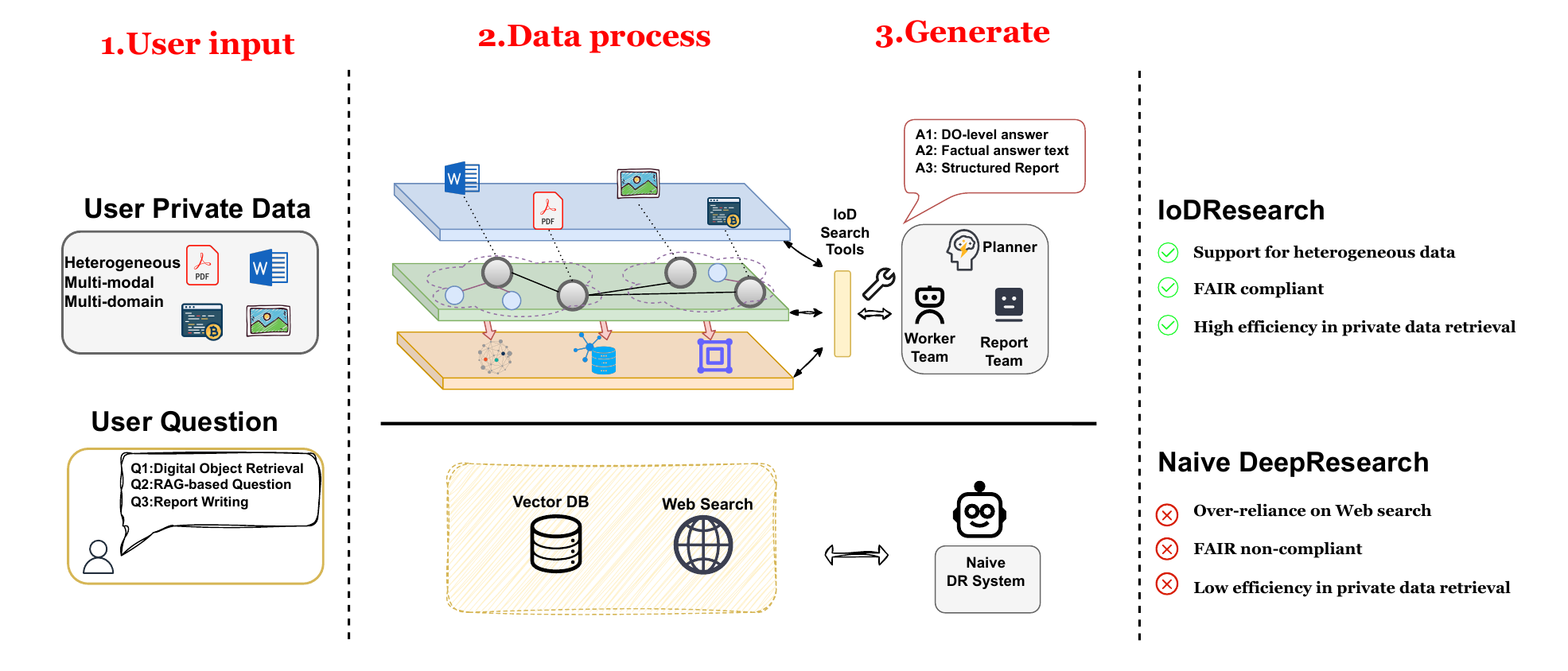}
    \caption{Motivation and architecture of IoDResearch, addressing naive Deep Research limitations via IoD-based multi-agent retrieval and reasoning.}
    \label{fig:zonglan}
\end{figure}

To address these challenges, we propose \emph{IoDResearch} (Internet of Data Research). As illustrated in Fig.~\ref{fig:zonglan}, IoDResearch serves both as a Deep Research framework for private, heterogeneous data under the IoD paradigm and as a novel data representation method within this paradigm. 
The framework constructs large-scale, multi-domain IoD networks \cite{shujukongjian, Meta-data-retrieval-for-data-infrastructure-via-RAG}, enabling standardized encapsulation and unified indexing of diverse data assets, thereby facilitating the comprehensive implementation of the FAIR principles. Moreover, a knowledge refinement layer is introduced to extract and refine latent knowledge.

Building upon this foundation, we further present \emph{IoDAgents}, a multi-agent framework designed to enhance data retrieval and improve the execution of Deep Research tasks. In addition, we introduce a Deep Research Benchmark under the IoD paradigm. Experimental results show that IoDResearch outperforms prior data representation methods and achieves superior performance across three tasks, validating both the feasibility and the potential of our approach.

The key contributions and advantages of our work are summarized as follows:  

\begin{itemize}  
\item \textbf{Private Data-Driven Deep Research via IoD.} 
We propose a private-data-driven Deep Research framework over heterogeneous multimodal data, built upon IoD. It enables unified encapsulation and indexing to fully realize the FAIR principles, while also providing a novel data representation paradigm for IoD.  

\item \textbf{IoD Heterogeneous Graph Representation.}  
We develop a three-layer IoD-based representation that transforms raw resources into FAIR-compliant digital objects(DO), and further refines them into atomic knowledge, vector embeddings, and knowledge graphs. This enables multi-granularity indexing and retrieval, from entire objects to fine-grained facts and semantic relations. 

\item \textbf{IoD DeepResearch Benchmark.} 
We introduce the first Deep Research benchmark under the IoD paradigm, which not only evaluates data representation capabilities in IoD settings but also assesses the unique ability of Deep Research systems to generate reports.  

\end{itemize}  

\section{Methodology}
\label{sec:format}
\subsection{Data preprocessing via IoD-based heterogeneous graph}
\begin{figure}[htbp]
    \centering
    \includegraphics[width=0.6\linewidth]{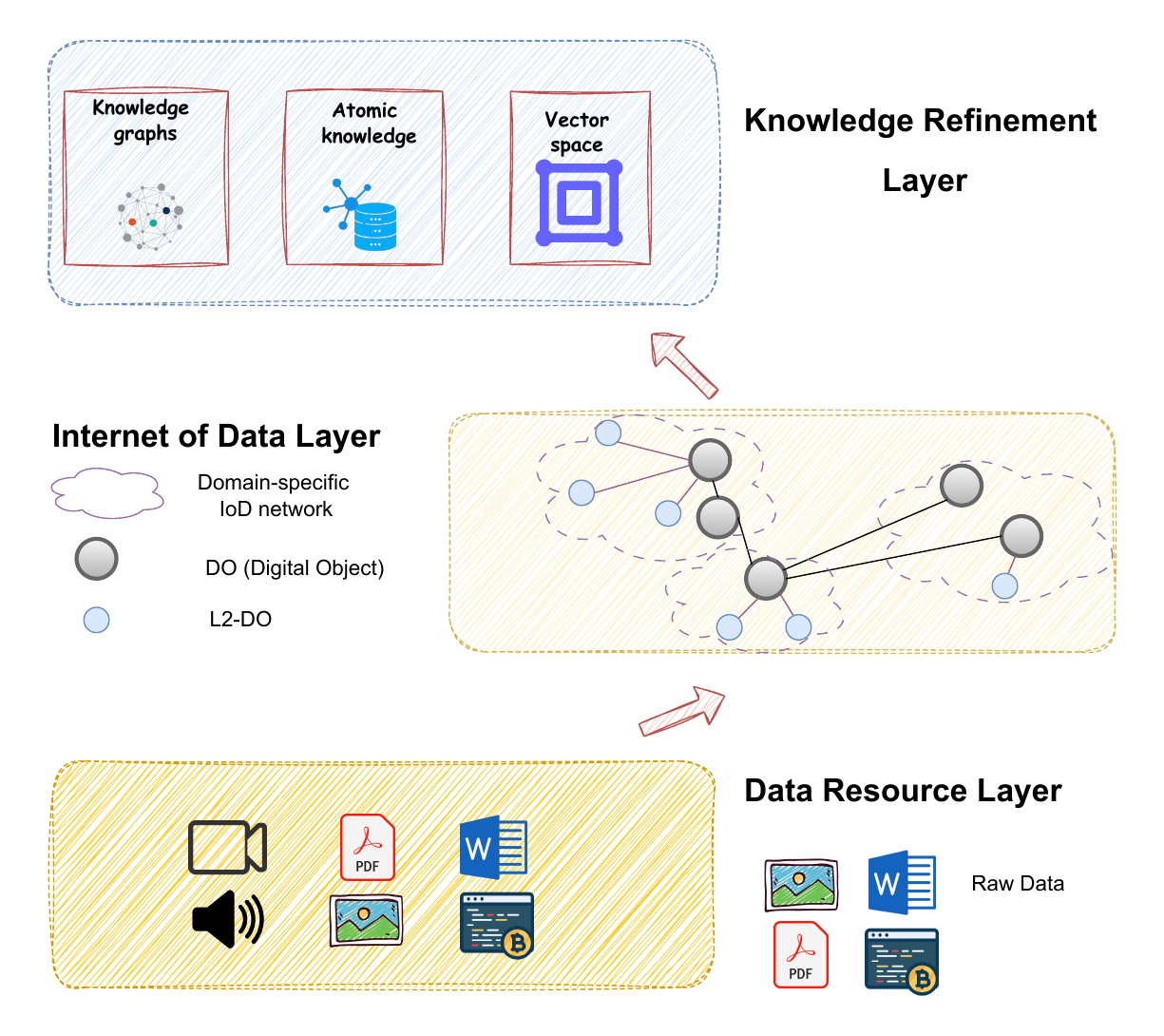}
    \caption{Transformation process from raw domain-specific resources to IoD-based heterogeneous graph representations.}
    \label{fig:从海量领域数据到数联网}
\end{figure}

To enable efficient data retrieval while adhering to the FAIR principles, we preprocess data through an IoD-based heterogeneous graph. As shown in Fig.~\ref{fig:从海量领域数据到数联网}, the overall architecture is organized into three layers:

\textbf{Data Resource Layer:} This layer encompasses multi-source, multimodal raw data, collected from diverse sources such as scientific publications, experimental datasets, etc.  

\textbf{Digital Object Layer:} At this layer, raw data entities are parsed, assigned a DOI, enriched with metadata, and encapsulated as digital objects. For long documents, each chunk is also encapsulated as a Level-2 Digital Object (L2-DO). Then all digital objects within a single domain form a domain-specific IoD. Following the IoD standard, multiple domain-specific IoDs are integrated into a global IoD network. 

\textbf{Knowledge Refinement Layer:} At this stage, the corpus is transformed into structured knowledge representations.
Vector representations enable efficient similarity-based retrieval across heterogeneous content.
Atomic knowledge captures fine-grained factual units, allowing precise identification of specific attributes or conclusions.
Knowledge graphs encode semantic structures and relational contexts, thereby supporting complex reasoning and multi-hop question answering.

\subsubsection{From data resources to the IoD}

\begin{figure}[htbp]
    \centering
    \includegraphics[width=1.0\linewidth]{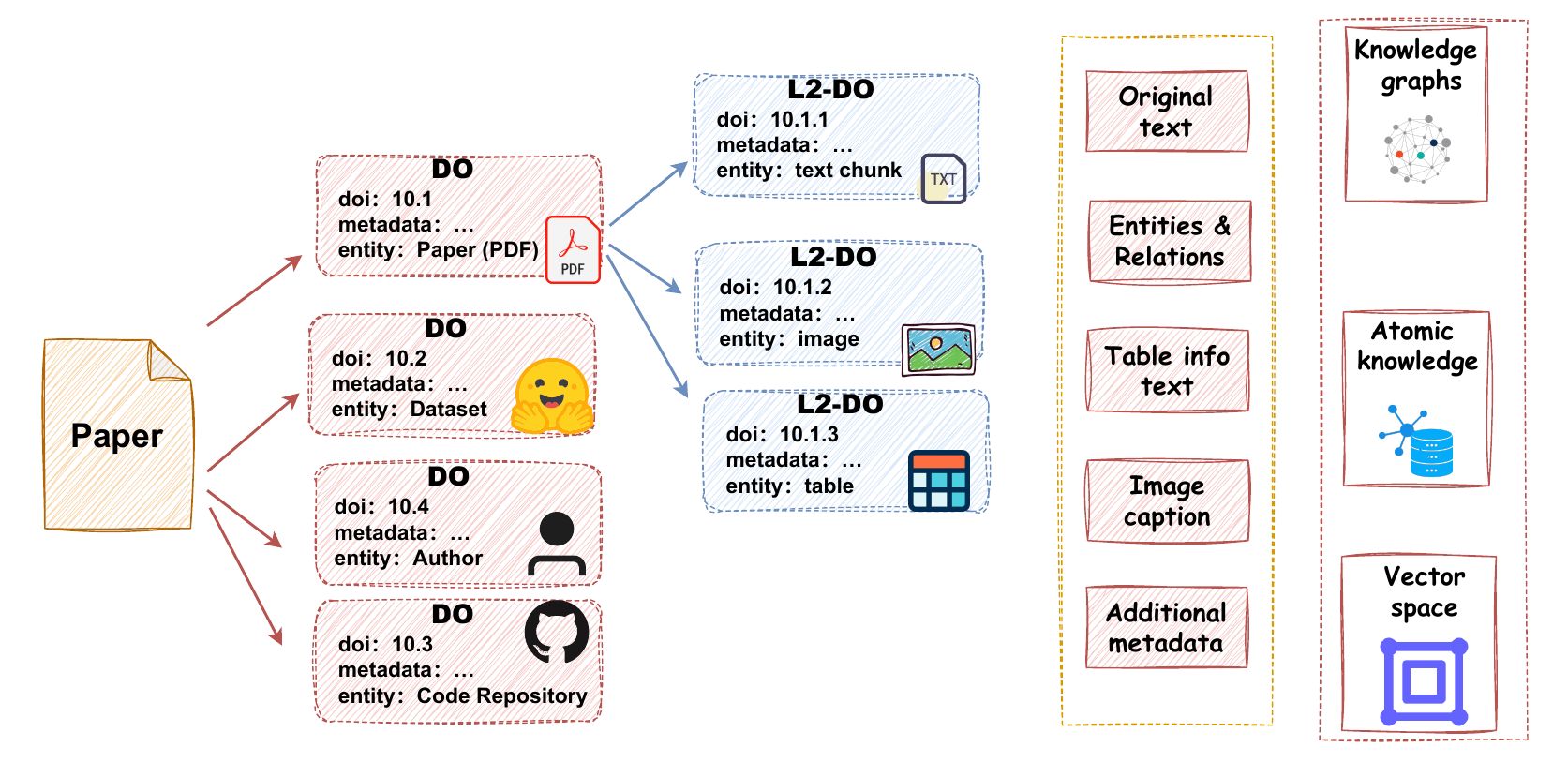}
    \caption{An example illustrating how a scientific paper and its associated resources are encapsulated into digital objects and further distilled into structured knowledge.}
    \label{fig:数据到数字对象到检索}
\end{figure}

For raw data, each entity is first assigned a unique identifier, enriched with metadata, and encapsulated as a digital object according to the DOA protocol \cite{shujukongjian}. The resulting digital objects are then integrated into the IoD, as illustrated in Fig.~\ref{fig:数据到数字对象到检索}.

\textbf{Entity Parsing and Multi-level Digital Objects.}
Entity files are parsed using open-source tools such as MinerU \cite{wang2024mineruopensourcesolutionprecise}. For long documents, each digital object is further divided into multiple chunks, with each chunk encapsulated as a Level-2 Digital Object (L2-DO) to enable fine-grained indexing and retrieval.

\textbf{Metadata Enrichment.}
Traditional IoD approaches mainly rely on manually annotated explicit metadata, which is often insufficient for uncovering implicit information and thereby limits retrieval efficiency. To address this limitation, we build upon conventional methods by enhancing explicit metadata with LLM-based automatic enrichment, extracting additional attributes such as content summaries, hypothetical questions, classification labels, and keywords. For non-textual entities such as audio and images, descriptive multimodal metadata is also generated. Furthermore, critical information identified during the knowledge refinement process is preserved as enriched metadata.


\subsubsection{Knowledge refinement} 
\textbf{Vector Representations.}
The content of digital objects, including entity text, metadata, textualized multimodal information, and textualized tabular data embedded into vector representations. During retrieval, these vectors support efficient similarity search, while the associated metadata provides reverse indexing back to the original digital objects.

\textbf{Knowledge Graph.}  
We leverage graph structures to enhance indexing and retrieval. Specifically, long documents are segmented into text passages; entities and relations are extracted from each passage using LLMs, forming graph nodes and edges. LLM profiling is then used to generate keywords and concise descriptions for both nodes and edges. Entities are typically indexed by their canonical names, while relations are enriched with multiple thematic keywords. Redundant entities and relations across passages are deduplicated and merged. The resulting knowledge graph thus consists of interconnected entities and relations optimized for retrieval and reasoning.  

\textbf{Atomic Knowledge.}  
Atomic knowledge refers to the minimal factual units distilled from digital objects, such as individual attributes, values, or statements that cannot be further decomposed. Unlike knowledge graphs, which emphasize the relational structure among entities, atomic knowledge captures fine-grained facts in an independent form (e.g., ``Ti\textsubscript{3}SiC\textsubscript{2}: melting\_point = 3200K'', ``Aspirin: typical\_dosage = 300\,mg''). These atomic units enable precise retrieval and direct fact matching, and can be flexibly combined during reasoning to support higher-level semantic inference. 

\subsection{IoD heterogeneous-graph-augmented retrieval}

\subsubsection{IoD search tools} 
All retrieval functions are encapsulated as tools via Model Context Protocol (MCP) and exposed to agents through standardized interfaces. We distinguish retrieval granularity from retrieval strategy. In terms of granularity, we support three tiers: digital-object retrieval, which returns relevant digital objects (DOs); content-chunk retrieval, which targets passages within a DO or the content of Level-2 Digital Objects (L2-DOs); and fine-grained retrieval at the knowledge-refinement layer, which returns atomic facts or subgraphs of the knowledge graph. Independently of granularity, retrieval modules can operate under multiple strategies—including keyword search, vector-based similarity, knowledge-graph-based reasoning, and multi-source/hybrid recall. To ensure quality, each retrieved item is accompanied by metadata (e.g., type, source, timestamp), enabling agents to automatically filter outdated, unreliable, or untrusted information. This design improves retrieval robustness and enhances both interpretability and trustworthiness in downstream reasoning.

\subsubsection{IoD agents for deep research} \label{Method:MultiAgentReasoning}
\begin{figure}[htbp]
    \centering
    \includegraphics[width=0.8\linewidth]{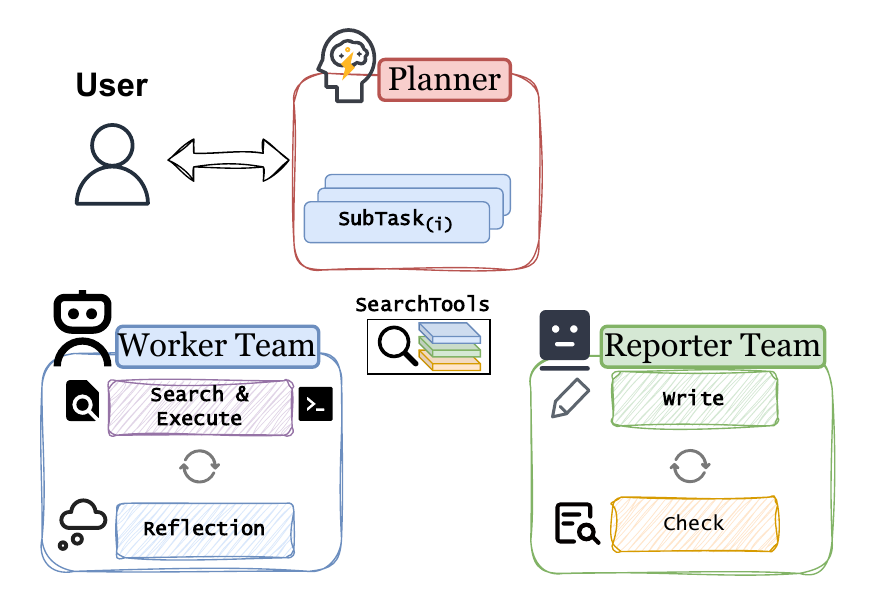}
    \caption{Multi-agent collaborative reasoning in IoDAgents, where the Planner decomposes user queries into subtasks, the Worker Team performs search and reflection, and the Reporter Team synthesizes and verifies the final report.}
    \label{fig:multiagent-reasoning}
\end{figure}

We design \textbf{IoDAgents}, a private-data-driven Deep Research system that supports a full spectrum of research activities, including efficient data retrieval, question answering, and the generation of long-form scientific reports. 
As illustrated in Fig.~\ref{fig:multiagent-reasoning}, the system consists of three functional teams: \emph{Planner}, \emph{Worker Team}, and \emph{Reporter Team}. A user first provides a question or a report topic to the Planner. The Planner generates a structured plan with specific steps, which is then returned to the user for confirmation or modification. If the query is ambiguous or lacks necessary details, the Planner proactively asks the user for clarification.  

Once the plan is confirmed, the Worker Team executes the tasks sequentially. For \emph{search tasks}, IoD-search-tools are invoked to retrieve information. The retrieved results are further filtered by LLMs to remove irrelevant content and, when necessary, summarized into key knowledge. If the retrieved context is insufficient, the query is refined and reissued iteratively. For execution tasks, tools (e.g., a code tool) are applied for data analysis, visualization, or chart generation.  

The \emph{Reporter Team} then integrates the results. The \emph{Writer Node} produces the final report or direct answer based on all retrieved knowledge, while the \emph{Checker Node} validates the output, identifies inconsistencies with the retrieved context, and applies corrections when needed.  

\section{Experiments}
\label{sec:pagestyle}
\subsection{Experimental setup}
\textbf{Dataset Construction} To enable systematic evaluation of Deep Research systems under the IoD paradigm, we present the \emph{IoD Deep Research Benchmark}. The benchmark is built upon a curated collection of over 500 high-quality documents or papers (approximately 6 million tokens) and associated resources from four representative domains—Chinese law, geophysical exploration, computer science, and molecular dynamics—serving as the raw data foundation. All resources were encapsulated into Digital Objects with enriched metadata in strict accordance with the IoD standard.

Based on this foundation, we manually designed three categories of tasks with the assistance of human experts and RAGAS\cite{ragas}:

\begin{itemize}
    \item Task 1: Digital Object Retrieval — 200 questions, reflecting a common user requirement in IoD scenarios.
    \item Task 2: RAG-based Question Answering — 800 questions, including 400 single-domain questions (single-hop:multi-hop = 1:1) and 400 cross-domain multi-hop questions.
    \item Task 3: Report Writing — 60 questions, consisting of 30 single-domain and 30 cross-domain questions, specifically designed to evaluate the unique ability of Deep Research systems to integrate complex knowledge and produce structured scientific writing.
\end{itemize}

In summary, the three task categories correspond to IoD-specific data representation (Task 1), the fundamental QA capability (Task 2), and the report-writing ability of agentic Deep Research systems (Task 3), together forming a comprehensive and hierarchical evaluation framework.

\textbf{Metrics}
For Task~1, we adopted conventional information retrieval metrics, namely Precision, Recall, and F1. 
For Task~2, evaluation was conducted using the metrics defined in RAGAS\cite{ragas} and its automated assessment framework. 
For Task~3, we employed both LLM-based and human expert assessments. For the LLM evaluation, we used Qwen-Turbo with temperature fixed at 0. Each report was independently scored three times, with the final score taken as the average. The LLM evaluated the reports along five dimensions: interest level, coherence and organization, relevance and focus, coverage, and breadth and depth. In parallel, human experts—kept single-blind to the identity of the method—applied the same five criteria while additionally assessing factual accuracy, thereby providing a more comprehensive and rigorous evaluation.

\textbf{Baseline}
We benchmarked the proposed IoDAgents system against several representative baselines. The comparison included: Naive RAG \cite{lewis2020retrieval}, a standard retrieval-augmented generation approach over a flat knowledge base; LightRAG\cite{LightRAG}, a lightweight graph-based RAG framework; DO-RAG \cite{Meta-data-retrieval-for-data-infrastructure-via-RAG}, an agentic RAG approach designed for IoD-based data representation; DeepSearcher \cite{deepsearcher2025}, a typical implementation of Local Deep Research that extends Deep Research with conventional RAG to retrieve private local data. All methods were evaluated on the same raw data foundation and with the same base LLM, Qwen-Turbo\cite{yang2025qwen3technicalreport}, to ensure fairness and reproducibility.

\subsection{Results and analysis}

\textbf{Task 1: Digital Object Retrieval.}  
As shown in Table~\ref{tab:task1}, IoDAgents achieves the best retrieval performance among all baselines, with an F1 score above 82\%. In IoD scenarios, digital-object retrieval is typically implemented through RAG-based approaches transplanted from text retrieval (e.g., DO-RAG). Compared to these approaches, our method delivers a clear improvement.  

This advantage arises from the IoD-compliant digital object representation, which integrates multilevel encapsulation, enriched metadata, and knowledge refinement. Such a design enables retrieval at multiple granularities, ranging from entire documents to fine-grained facts and semantic relations. Consequently, IoDAgents achieves more reliable and accurate retrieval in heterogeneous data environments.

\begin{table}[ht]
\centering
\caption{Performance on Task~1 (Retrieval).}
\resizebox{0.6\linewidth}{!}{
\begin{tabular}{l|ccc}
\hline
\textbf{Method} & Precision & Recall & F1 \\
\hline
NaiveRAG\cite{lewis2020retrieval}    
& 55.22 & 70.82 & 62.05 \\
DO-RAG\cite{Meta-data-retrieval-for-data-infrastructure-via-RAG} 
& 69.51 & 84.34 & 76.21 \\
LightRAG\cite{LightRAG}              
& 73.15 & 85.69 & 78.93 \\
\textbf{IoDResearch}                 
& 76.26 & 90.18 & 82.64 \\
\hline
\end{tabular}}
\label{tab:task1}
\end{table}

\textbf{Task 2: Question Answering.}  
For QA tasks (Table~\ref{tab:task2}), IoDAgents consistently outperforms all baselines in both single-domain and cross-domain settings. Cross-domain queries, especially those involving multi-hop reasoning, pose particular challenges for conventional RAG systems. Our method addresses these challenges in two ways: first, the IoD-compliant digital object representation provides more accurate and domain-aware retrieval across heterogeneous sources; second, the multi-agent framework enables iterative planning, retrieval, and self-checking, which is particularly effective for multi-hop questions.

\begin{table}[ht]
\centering
\caption{Performance on Task~2 (QA).}
\resizebox{\linewidth}{!}{
\begin{tabular}{l|cccccc}
\hline
\textbf{Method (Single-domain)} 
& Ans. Acc. & Ans. Faith. & Ans. Rel. & Ctx. Prec. & Ctx. Rec. & Ctx. F1 \\
\hline
NaiveRAG\cite{lewis2020retrieval}   
& 70.28 & 84.53 & 87.52 & 60.39 & 75.77 & 67.21 \\
DO-RAG\cite{Meta-data-retrieval-for-data-infrastructure-via-RAG}  
& 71.45 & 84.12 & 87.34 & 61.35 & 76.45 & 68.07 \\
LightRAG\cite{LightRAG}               
& 75.55 & 85.10 & 86.59 & 64.89 & 75.95 & 69.98 \\
\textbf{IoDResearch}                  
& 79.98 & 87.33 & 90.20 & 65.35 & 80.45 & 72.11 \\
\hline
\textbf{Method (Cross-domain)} 
& Ans. Acc. & Ans. Faith. & Ans. Rel. & Ctx. Prec. & Ctx. Rec. & Ctx. F1 \\
\hline
NaiveRAG\cite{lewis2020retrieval}   
& 42.42 & 69.98 & 66.05 & 44.38 & 45.00 & 44.69 \\
DO-RAG\cite{Meta-data-retrieval-for-data-infrastructure-via-RAG}  
& 50.32 & 70.31 & 66.58 & 46.52 & 48.65 & 47.56 \\
LightRAG\cite{LightRAG}               
& 56.67 & 75.59 & 76.91 & 50.02 & 52.50 & 51.23 \\
\textbf{IoDResearch}                  
& 59.40 & 78.07 & 78.18 & 52.02 & 53.50 & 52.75 \\
\hline
\end{tabular}}
\label{tab:task2}
\end{table}

\textbf{Task 3: Scientific Report Generation.}  
Table~\ref{tab:task3} presents the results of report generation. Human experts rated our system above 7.0 in single-domain and 6.4 in cross-domain scenarios, clearly outperforming DeepSearcher and other baselines. These results suggest that existing Deep Research approaches, exemplified by DeepSearcher, remain insufficient for handling private data, as they largely depend on Naive RAG strategies. Likewise, conventional RAG pipelines struggle to generate high-quality reports directly.
We also conduct an ablation by removing the multi-agent component (\emph{IoDResearch without Agent}). This variant yields consistently lower scores than the full system, confirming that agent-based collaboration is crucial  in report generation. Overall, the results highlight that IoDResearch, with robust private-data support and multi-agent design, effectively bridges the gap left by prior methods.

\begin{table}[ht]
\centering
\caption{Performance on Task~3 (Report writing)}
\resizebox{\linewidth}{!}{
\begin{tabular}{l|cc|cc}
\hline
\textbf{Method} & \multicolumn{2}{c|}{\textbf{LLM-as-Judge Score}} & \multicolumn{2}{c}{\textbf{Human Expert Score}} \\
\cline{2-5}
 & Single-domain & Cross-domain & Single-domain & Cross-domain \\
\hline
Zero-shot LLM & 7.61 & 7.45 & 5.65 & 5.23 \\
Light RAG\cite{LightRAG} & 7.95 & 7.86 & 6.53 & 5.88 \\
IoDResearch without Agent & 8.03 & 7.92 & 6.56 & 5.94 \\
DeepSearcher\cite{deepsearcher2025} & 8.13 & 8.08 & 6.77 & 6.02 \\
\textbf{IoDResearch} & 8.31 & 8.23 & 7.01 & 6.45 \\
\hline
\end{tabular}}
\label{tab:task3}
\end{table}

\textbf{Overall Analysis.}
Across all three tasks, IoDResearch demonstrates consistent and significant improvements over existing baselines. The IoD-compliant digital object representation enhances retrieval accuracy by enabling multi-granularity access to heterogeneous data, while the multi-agent framework improves reasoning reliability and report quality. These advantages are reflected not only in conventional IR metrics and QA performance but also in holistic report evaluation by both LLM-as-Judge and human experts. The results collectively validate that IoDResearch effectively bridges the gap between FAIR-compliant data representation and practical Deep Research applications, particularly in scenarios involving private or domain-specific data where conventional RAG pipelines fall short.

\section{Limitations and Future Work}

\label{sec:typestyle}
As the current pipeline is primarily text-based, information loss may occur during preprocessing and context construction for multimodal data. In future work, we will explore an end-to-end multimodal pipeline using multimodal embeddings and LLMs, and investigate post-training strategies to better assess the upper-bound performance of the proposed framework.

\section{Acknowledgment}
This work was supported by the National Natural Science Foundation of China under Grant No.~62595734.

\section{Conclusion}

\label{sec:typestyle}
In this work, we introduced IoDResearch, a private data-centric Deep Research framework that operationalizes the IoD paradigm. By encapsulating heterogeneous assets into FAIR-compliant digital objects and refining them into atomic knowledge and knowledge graphs, IoDResearch enables high-recall and complex reasoning over private data sources. Together with a multi-agent system, it supports not only reliable question answering but also structured scientific report generation. Furthermore, we released the IoD DeepResearch Benchmark, which provides the first systematic evaluation of data representation and agentic Deep Research capabilities under the IoD paradigm. Experimental results show that IoDResearch outperforms the selected RAG and Deep Research baselines on the proposed IoD DeepResearch Benchmark across the evaluated scenarios and task settings.
\bibliographystyle{unsrt}  
\bibliography{references}

\end{document}